\newif\ifColor\Colortrue \Colorfalse
\Colortrue
\PassOptionsToPackage{super,comma,numbers,sort&compress}{natbib}  
\documentclass[preprint, 5p, 8pt, lefttitle]{elsarticle}

\usepackage{xpatch}
\xpatchcmd{\MaketitleBox}{\hrule\vskip12pt}{\vspace{-2\baselineskip}}{}{}
\xpatchcmd{\MaketitleBox}{\hrule}{}{}{}

\usepackage[switch, left]{lineno}
\modulolinenumbers[1]
\usepackage[utf8]{inputenc}
\usepackage{amssymb}
\usepackage{amsmath}
\usepackage{amsfonts}
\usepackage{upgreek}
\usepackage{gensymb}
\usepackage{setspace}
\usepackage{booktabs}
\usepackage[english]{babel}
\usepackage{mathtools}
\usepackage{mhchem}
\usepackage{hyperref}
\usepackage[final]{microtype}
\usepackage{csquotes,lmodern}
\usepackage{subcaption}
\usepackage{etoolbox}
\usepackage{color}
\usepackage{multibib}
\newcites{methods}{References}
\usepackage[labelfont=bf]{caption}

\renewcommand{\figurename}{Fig.}
\usepackage{sidecap}
\usepackage{adjustbox}

\makeatletter
\renewenvironment{abstract}{\global\setbox\absbox=\vbox\bgroup
  \hsize=\textwidth\def\baselinestretch{0}%
 \par\unskip\noindent\unskip\ignorespaces}
 {\egroup}

\def\keyword{%
  \def\sep{\unskip, }%
 \def\MSC{\@ifnextchar[{\@MSC}{\@MSC[2000]}}
  \def\@MSC[##1]{\par\leavevmode\hbox {\it ##1~MSC:\space}}%
 \def\PACS{\par\leavevmode\hbox {\it PACS:\space}}%
  \def\JEL{\par\leavevmode\hbox {\it JEL:\space}}%
 \global\setbox\keybox=\vbox\bgroup\hsize=\textwidth
  \normalsize\normalfont\def\baselinestretch{0}
 \parskip\z@
  \noindent\textit{Some important words: }   <--- Edit as necessary
  \raggedright                         
  \ignorespaces}

\def\ps@pprintTitle{%
     \let\@oddhead\@empty
     \let\@evenhead\@empty
     \def\@oddfoot{\footnotesize\itshape
      \ifx\@journal\@empty  
       \else\@journal\fi\hfill\today}%
     \let\@evenfoot\@oddfoot}
     
\makeatother

\begin{document}
\begin{frontmatter}

\title{\doublespacing\textbf{\fontsize{20}{20}\selectfont{\textsf{High thermoelectric performance in metallic NiAu alloys}}}}

\author[1]{\textsf{\small\textbf{F. Garmroudi}}\corref{cor1}}
\author[1]{\textsf{\small\textbf{M. Parzer}}\corref{cor1}}
\author[1]{\textsf{\small\textbf{A. Riss}}}
\author[3]{\textsf{\small\textbf{C. Bourgès}}}
\author[2]{\textsf{\small\textbf{S. Khmelevskyi}}}
\author[4,5]{\textsf{\small\textbf{T. Mori}}}
\author[1]{\textsf{\small\textbf{E. Bauer}}}
\author[1]{\textsf{\small\textbf{A. Pustogow}}\corref{cor1}}

\cortext[cor1]{fabian.garmroudi@tuwien.ac.at, michael.parzer@tuwien.ac.at, pustogow@ifp.tuwien.ac.at}
%
%
%
\address[1]{\textsf{Institute of Solid State Physics, TU Wien, 1040 Vienna, Austria}}
\address[2]{\textsf{International Center for Young Scientists (ICYS), National Institute for Materials Science, Tsukuba, Japan}}
\address[3]{\textsf{Research Center for the Computational Materials Science and Engineering, TU Wien, 1040 Vienna, Austria}}
\address[4]{\textsf{International Center for Materials Nanoarchitectonics (WPI-MANA), National Institute for Materials Science, Tsukuba, Japan}}
\address[5]{\textsf{Graduate School of Pure and Applied Sciences, University of Tsukuba, Tsukuba, Japan}}
\selectlanguage{english}

\begin{abstract}
\noindent \textsf{\fontsize{10}{10}\selectfont{Thermoelectric (TE) materials seamlessly convert thermal into electrical energy and vice versa, making them promising for applications such as power generation or cooling. Although historically the TE effect was first discovered in metals, state-of-the-art research mainly focuses on doped semiconductors with large figure of merit, \textit{zT}, that determines the conversion efficiency of TE devices. While metallic alloys have superior functional properties, such as high ductility and mechanical strength, 
they have mostly been discarded from investigation in the past due to their small Seebeck effect. 
Here, we realize unprecedented TE performance in metals by tuning the energy-dependent electronic scattering. Based on our theoretical predictions, we identify binary NiAu alloys as promising candidate materials and experimentally discover colossal power factors up to 34\,mWm\mbox{\boldmath$^{-1}$}K\mbox{\boldmath$^{-2}$} (on average 30\,mWm\mbox{\boldmath$^{-1}$}K\mbox{\boldmath$^{-2}$} from 300 to 1100\,K), which is more than twice larger than in any known bulk material above room temperature. This system reaches a \textit{zT} up to 0.5, setting a new world record value for metals. NiAu alloys are not only orders of magnitude more conductive than heavily doped semiconductors, but also have large Seebeck coefficients originating from an inherently different physical mechanism: within the Au \textit{s} band conduction electrons are highly mobile while holes are scattered into more localized Ni \textit{d} states, yielding a strongly energy-dependent carrier mobility. Our work challenges the common belief that good metals are bad thermoelectrics and presents an auspicious paradigm for achieving high TE performance in metallic alloys through engineering electron-hole selective \textit{s--d} scattering.}}\end{abstract}

\end{frontmatter}


Thermoelectrics are promising materials for various applications, such as thermometry, refrigeration and power generation \cite{hendricks2022keynote}. Following Ioffe's proposal in the 1930s \cite{ioffe1932problem}, the majority of investigations on TE materials focused on semiconductors, with the general idea being that a band gap is required to obtain a strong asymmetry in the electronic density of states (DOS) $N(E)$ around the Fermi energy $E_\text{F}$, prerequisite for a large Seebeck effect. This predominant notion has confined research to a narrow fraction of known materials, neglecting metallic systems. Despite remarkable progress achieved by adhering to this paradigm, the most substantial improvement of the dimensionless figure of merit $zT=(S^2 \sigma/\kappa)T$ has been obtained through the reduction of thermal conductivity $\kappa$ via phonon engineering \cite{hochbaum2008enhanced,
biswas2012high,zhao2014ultralow,kim2015dense}, which is inherently limited. Therefore, identifying new enhancement principles for the power factor, $PF=S^2 \sigma$, is of great interest, especially for power generation \cite{liu2015n,liu2016importance} or the emerging field of active cooling \cite{zebarjadi2015electronic,adams2019active,komatsu2021macroscopic}. Here, $S$ and $\sigma$ are the Seebeck coefficient and electrical conductivity. Usually, these two electronic transport properties are entangled such that an increase of $\sigma$ also decreases $S$ and vice versa. 
While numerous strategies to tackle this problem and enhance the power factor have been proposed \cite{heremans2008enhancement,
pei2011convergence,ahmed2017thermoelectric,tsujii2019observation}, the simultaneous realization of a high $S$ and $\sigma$ remains one of the toughest challenges for TE research.

\begin{figure*}[t]
\newcommand{\setwidth}{0.45}
			\centering
			\hspace*{0cm}
		\includegraphics[width=1\textwidth]{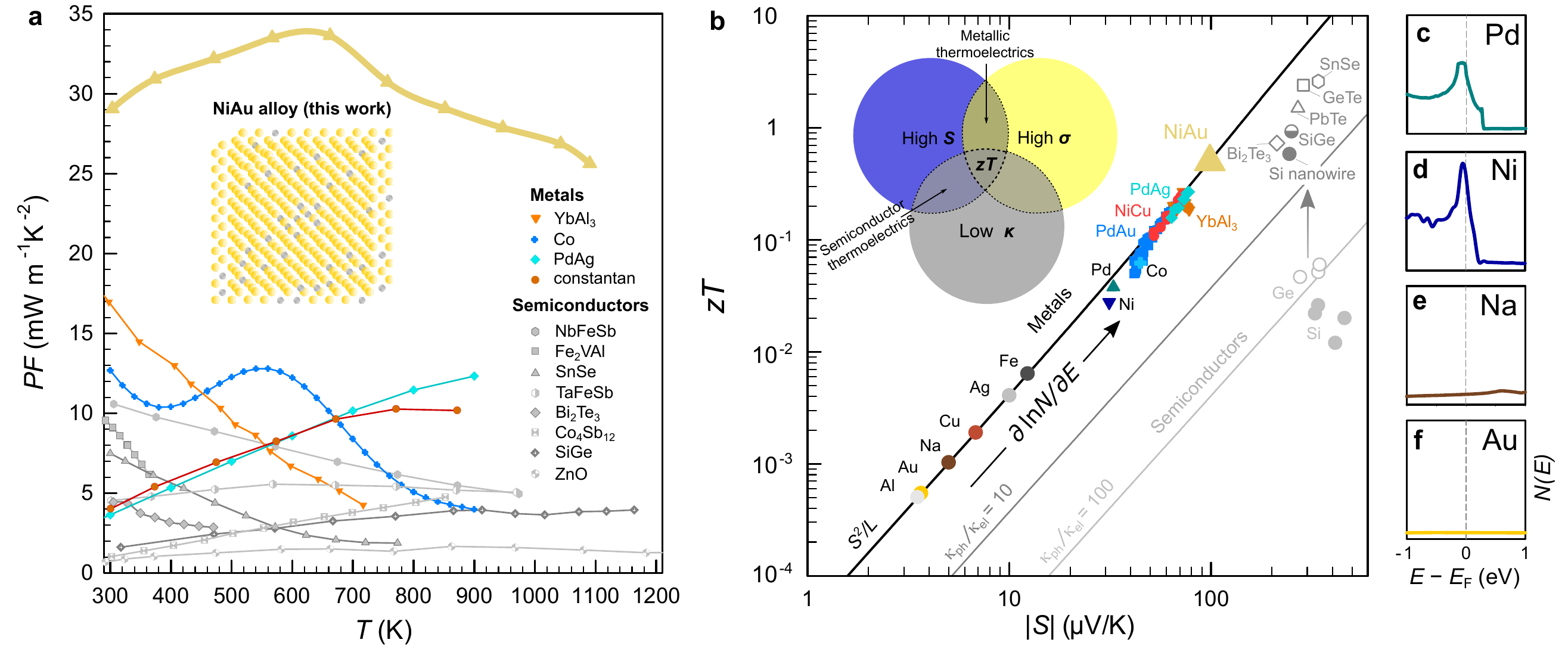}
	\caption{\textbf{$\vert$ Thermoelectric performance of NiAu alloys compared to today's best thermoelectrics.} \textbf{a}, Power factor of \ce{Ni_{0.1}Au_{0.9}} and various semiconducting \cite{ohtaki1996high,wang2008enhanced,
shi2011multiple,he2016achieving,
cha2019ultrahigh,zhu2019discovery,garmroudi2021boosting,qin2021power} and metallic systems \cite{rowe2002electrical,ho1993thermoelectric,
mao2015high,watzman2016magnon}. \textbf{b}, Wiedemann-Franz plot of thermoelectrics: universal scaling $zT\propto S^2/L$ defines the TE performance in metals. Conventional semiconductors like Si \cite{stranz2013thermoelectric} and Ge \cite{ohishi2016thermoelectric} are orders of magnitude below $S^2/L$ as the lattice thermal conductivity is much larger than the electronic contribution $\kappa_\text{ph}\gg\kappa_\text{el}$. Despite extremely low $\kappa_\text{ph}$, state-of-the-art TE semiconductors \cite{hochbaum2008enhanced,joshi2008enhanced,pei2011high,
zhao2014ultralow,li2018low,witting2019thermoelectric} still have several times lower $zT$ compared to the \textit{metallic limit} $S^2/L$, where TE performance is optimal for a given value of $S$. \textbf{c}-\textbf{f}, Density of states $N(E)$ for metallic elements with delocalized (e,f) $s$ and more localized (c,d) $d$ states near $E_\text{F}$. Since $s$-$d$ scattering is proportional to $N(E)$, the logarithmic derivative $d\ln{N(E)}/dE$ determines $zT\propto S^2$ in metals.
} 
	\label{Fig1}
\end{figure*}

\begin{figure*}[t]
\newcommand{\setwidth}{0.45}
			\centering
			\hspace*{0cm}
		\includegraphics[width=0.97\textwidth]{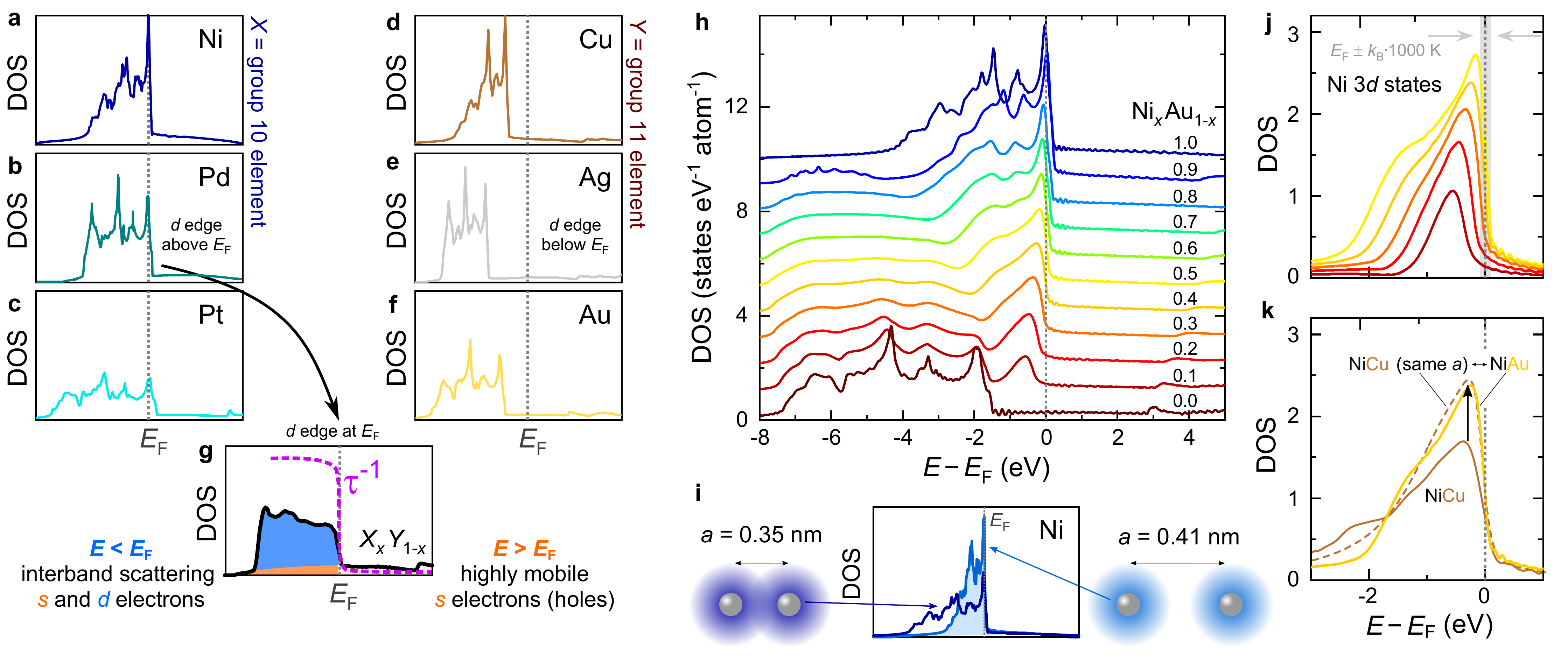}
	\caption{\textbf{$\vert$ Electron-hole selective \textit{s}--\textit{d} scattering in binary metallic alloys.} \textbf{a}-\textbf{f}, Electronic DOS $N(E)$ of group 10 ($X$) and 11 ($Y$) elements. While for $Y$ the filled $d$ states lie more than 1~eV below $E_{\rm F}$, the partial $d$ occupation in $X$ yields large DOS at $E_{\rm F}$; paramagnetic $N(E)$ shown for $X=$~Ni. \textbf{g}, The steep slope of $N(E_{\rm F})$ involves strong $s$--$d$ scattering at $E<E_{\rm F}$, resulting in a steep gradient of the scattering rate $\tau^{-1}$ across $E_{\rm F}$. \textbf{h}, In binary alloys $X_{x}Y_{1-x}$, such as \ce{Ni_{x}Au_{1-x}}, the $d$ edge of $X$ atoms is successively tuned through $E_{\rm F}$ upon changing $x$. \textbf{i}, Increasing the lattice spacing of Ni atoms to $a_{\rm Au}=0.41$~nm reduces the bandwidth and steepens the $d$ edge slope compared to elemental Ni ($a_{\rm Ni}=0.35$~nm). \textbf{j}, Partial DOS of Ni $d$ states in \ce{Ni_{x}Au_{1-x}}; grey bar indicates thermal energy of 1000\,K around $E_{\rm F}$. \textbf{k}, \textit{Ab initio} calculations yield similar $N_d(E)$ for NiCu and NiAu when the same lattice parameter is used, identifying bandwidth tuning as the pivotal factor.
	} 
	\label{Fig2}
\end{figure*}
\begin{figure}[t]
\newcommand{\setwidth}{0.45}
			\centering
			\hspace*{-0.3cm}
		\includegraphics[width=0.5\textwidth]{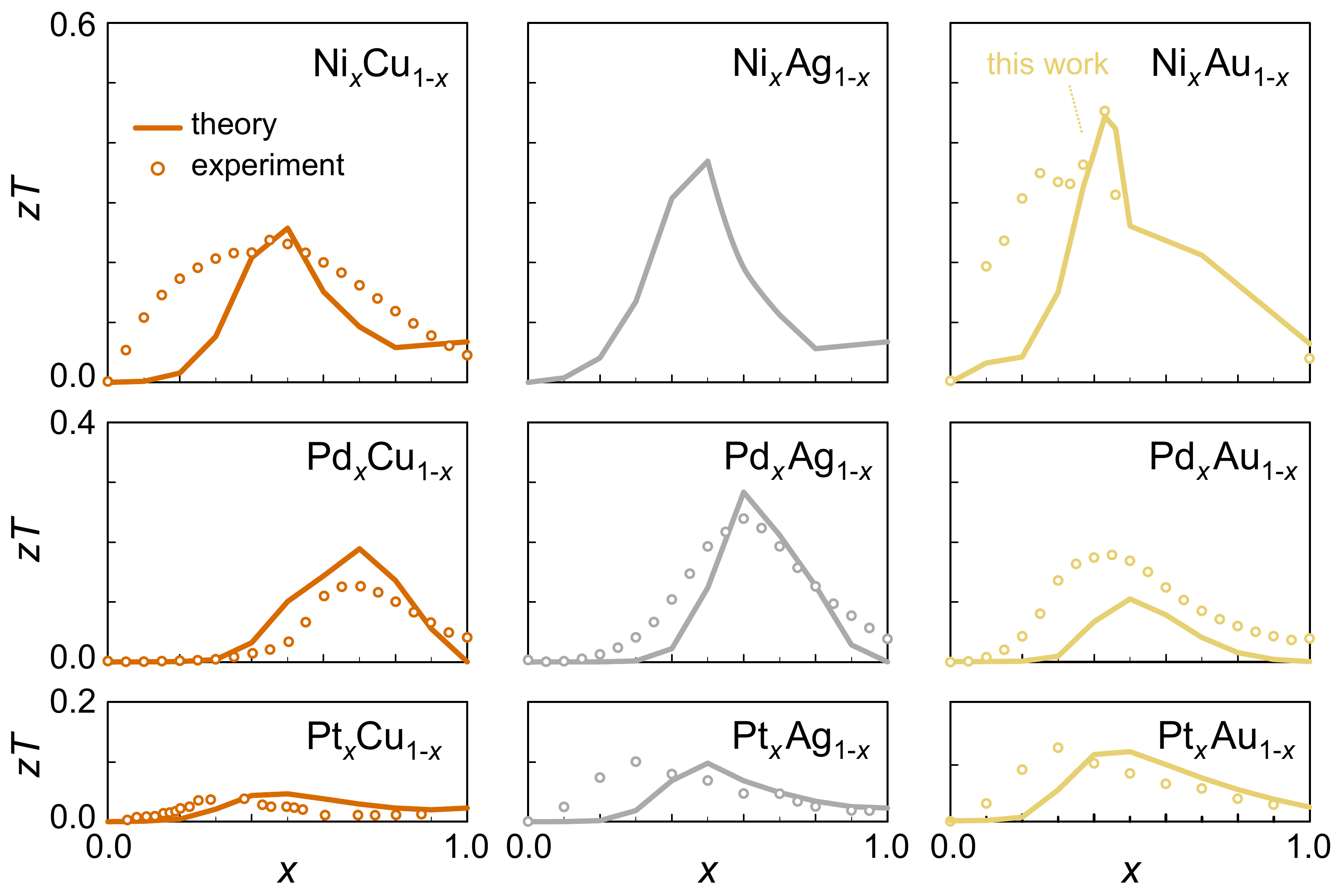}
	\caption{\textbf{$\vert$ \textit{zT} from \textit{s}--\textit{d} scattering compared to experiment.} The composition-dependent $zT$ at $1000\,$K, calculated within the theoretical framework of \textit{s}--\textit{d} scattering $\tau^{-1}\propto N(E)$, agrees well with experimental results on $X_{x}Y_{1-x}$ ($X=$~Ni, Pd, Pt; $Y=$~Cu, Ag, Au) from literature \cite{rudnitskii1956thermoelectric,
ho1978thermal,ho1983electrical,ho1993thermoelectric} and this work. Highest $zT$ is reached for Ni alloys due to the more localized 3$d$ states compared to the 4$d$ and 5$d$ bands of Pd and Pt. Among group 11 elements, Au alloys show the highest performance due to narrower bandwidth of Ni $d$ states. 
		} 
	\label{Fig3}
\end{figure}

Here, we report ultrahigh power factors $>$ 30\,mWm$^{-1}$K$^{-2}$ in binary NiAu metallic alloys, vastly exceeding those of any known bulk material above room temperature (see Fig.\,\ref{Fig1}a) as well as $zT$ values up to 0.5, higher than in any other metallic system (see Fig.\,\ref{Fig1}b). Our electronic structure calculations reveal that this is the result of a strongly energy-dependent scattering rate $\tau^{-1}$ arising from a steep gradient of Ni $d$ states around $E_{\rm F}$. While $s$-type holes are scattered into more localized Ni $d$ states below $E_{\rm F}$, the mobility remains high for conduction electrons above $E_{\rm F}$. 
Furthermore, we emphasize that in metals where thermal transport is largely dictated by the electron contribution, only the Seebeck coefficient has to be tuned in order to enhance the figure of merit, which simplifies to $zT=S^2/L$ due to the Wiedemann-Franz law (as illustrated in Fig.\,\ref{Fig1}b), where $L$ represents the Lorenz number. Hence, our work constitutes a promising new paradigm in TE research, circumventing the multidimensional optimization problem in semiconductors that involves balancing the trade-offs between $S$, $\sigma$ and $\kappa$. 

In the following, we will demonstrate that a band gap in the electronic structure is not necessary for designing high-performance thermoelectrics. Fig.\,\ref{Fig1}c-f shows the DOS near the Fermi energy for several metallic elements. Alkali metals like Na, as well as transition metals like Cu, Ag and Au, have a symmetric distribution $N(E)$ around $E_\text{F}$ due to their partly filled $s$ orbitals with a large bandwidth $W$. In these systems, $zT$ is orders of magnitude smaller compared to transition metals with sharp, more localized $d$ states near $E_\text{F}$ like Ni or Pd (Fig.\,\ref{Fig1}b). We will show that this increased TE performance originates from the energy dependence of $\tau^{-1}$ upon $s \rightarrow d$ scattering and quantitatively estimate $zT$ directly from $N(E)$. To examine this relationship for the different elements, we constructed a \textit{periodic table of densities of states} (see Extended Data Fig.\,1) which summarizes the electronic structures of solid elements and can be used as a starting point for future studies searching for high TE performance in metals. 

The concept of $s$--$d$ scattering dates back to Mott's work in 1935 describing the electronic structure of the binary Pd-Ag system \cite{mott1935discussion}. In these alloys, high-velocity Ag $s$-type carriers scatter into rather immobile Pd $d$ states, which increases the residual resistivity and also the diffusion term of the Seebeck coefficient. Similarly, an unusually high Seebeck effect is found in isovalent NiCu alloys, exploited in thermocouples. Since then, however, metals have been largely overlooked by the majority of the TE community, dismissing $s$--$d$ scattering as a potential route to design new thermoelectrics. In our quest for high-performance metallic thermoelectrics, we investigated all combinations of binary alloys between group 10 and 11 transition metals and estimated the TE performance of these systems above room temperature based on the evolution of their electronic structure. Fig.\,\ref{Fig2}a-f shows $N(E)$ of $X=\text{Ni, Pd, Pt}$ and $Y= \text{Cu, Ag, Au}$ in their pure form. The $d$ levels are not fully occupied for $X$ and, in the case of Ni, $N(E_{\rm F})$ surpasses the Stoner criterion $JN(E_\text{F})>1$ ($J$ is the exchange interaction constant), giving rise to ferromagnetism below $T_\text{C}=633\,$K. For $Y$, the $d$ states are completely filled and lie more than 1 eV below $E_\text{F}$. The small featureless DOS of high-velocity $s$ and $p$ electrons around $E_\text{F}$ makes Cu, Ag and Au the best metallic conductors. In the solid solution $X_{x}Y_{1-x}$, the $d$ edge of the $X$ atoms is shifted across $E_\text{F}$ with decreasing $x$.
Our \textit{ab initio} calculations on Ni$_{x}$Au$_{1-x}$ demonstrate the successive pile-up of Ni $d$ states below $E_\text{F}$ (Fig.\,\ref{Fig2}h,j), which is similar for all other combinations $X_{x}Y_{1-x}$ as shown in Extended Data Fig.\,2. 

Assessing transport properties from the DOS, one would usually expect a sizeable \textit{positive} Seebeck coefficient from the resulting \textit{negative} slope of $N$ at $E_\text{F}$ (see Fig.\ref{Fig2}g-k) related to the first term of the well-known Mott formula
\begin{equation}
\label{Mottformula}
S = -\frac{\pi^2 k_\text{B}^2}{3e}T\left(\frac{\partial\,\ln{N_\text{c}}}{\partial E}-\frac{\partial\,\ln{\tau^{-1}}}{\partial E}\right)_{E = E_\text{F}}.
\end{equation}
Here, $N_\text{c}$ represents the DOS of mobile conduction electrons, yielding negligibly small $\partial \ln N_\text{c}/\partial E$ in $X_xY_{1-x}$ metallic alloys due to the large bandwidth of $s$ and $p$ states. The second term in Eq.\,\ref{Mottformula} arises from the energy-dependent scattering rate $\tau^{-1}$. While this mechanism is commonly neglected in semiconductors, TE transport in such metallic alloys is dominated by $\partial \ln \tau^{-1}/\partial E$. The more localized Ni $d$ states hardly contribute to $N_\text{c}$ \cite{mott1935discussion} but induce strong inter-orbital scattering. According to Fermi's golden rule, the phase space of $s$--$d$ scattering scales with $N_d$, which leads to a pronounced energy dependence of $\tau^{-1}\propto N_d$ as indicated by the dashed magenta line in Fig.\,\ref{Fig2}g. Consequently, hole-type carriers at $E<E_\text{F}$ contribute much less to charge transport than conduction electrons above $E_\text{F}$, yielding a large \textit{negative} $S \propto \,\partial\,\ln{N_d}/\partial E$ in these metallic alloys.  

Based on the DOS we estimate in Fig.\,\ref{Fig3} the figure of merit of \ce{$X$_{x}$Y$_{1-x}} via $zT\propto S^2 \propto \left(\partial \text{ln}N_d/\partial E\right)^2$ and obtain remarkable agreement with the experimental results obtained here and in literature \cite{rudnitskii1956thermoelectric,
ho1978thermal,ho1983electrical,ho1993thermoelectric}. Note that the calculated TE properties already include the deteriorating effect of inelastic phonon scattering, becoming prominent at $T> 300$\,K, via the Nordheim-Gorther rule (Methods). 
On the one hand, from Fig.\,\ref{Fig3} it becomes clear that Ni is the most promising group 10 element for the design of TE metals, as it has the highest $zT$ due to more localized $3d$ states (see Fig.\,\ref{Fig2}a-c) and also lower cost than Pd and Pt. On the other hand, among Ni alloys with group 11 elements, Ag and Au exhibit the highest $zT$, which we attribute to the larger lattice parameter compared to Cu (no experimental data exist for Ni-Ag due to the immiscibility of this system). As a consequence of this negative chemical pressure there is less overlap between Ni $d$ orbitals, leading to even more localized states with smaller bandwidth, and hence, larger $S \propto \,\partial\,\ln{N_d}/\partial E$ (see Fig.\ref{Fig2}i,j). 

\begin{figure*}[h!]
 \newcommand{\setwidth}{0.45}
			\centering
			\hspace*{-0.1cm}	
			\vspace*{0cm}
 \begin{minipage}[t]{0.67\textwidth}
    \adjincludegraphics[width=0.95\textwidth,valign=T]{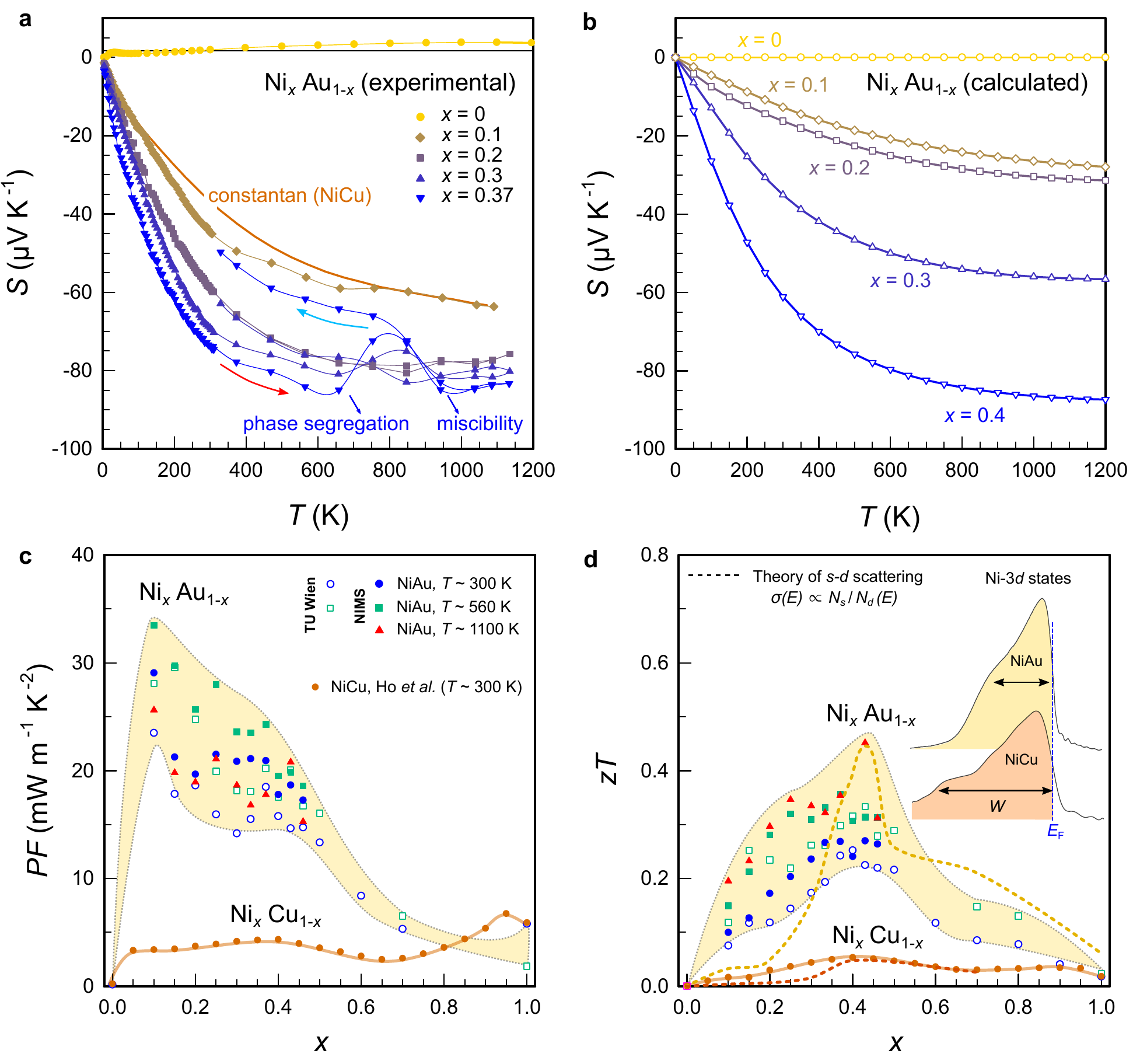}
  \end{minipage}\hfill
  \begin{minipage}[t]{0.32\textwidth}
    \caption{\textbf{$\vert$ Thermoelectric properties of the Ni-Au system.} \textbf{a}, Temperature-dependent Seebeck coefficient of selected Au-rich \ce{Ni_{x}Au_{1-x}} alloys. The Seebeck coefficient of constantan (\ce{Ni_{0.45}Cu_{0.55}}) is shown for reference. Au-rich samples with $x<0.3$ are (meta)stable and show no hysteresis even when measuring up to 1100\,K. With increasing Ni concentration, however, phase segregation occurs during the measurement, followed by reentrant miscibility above the critical temperature of the solid solution. \textbf{b}, Temperature-dependent Seebeck coefficient calculated from electronic DOS within the framework of \textit{s--d} scattering. \textbf{c}, Composition-dependent power factor of the Ni-Au system at various temperatures within the single-phase regime. Data have been obtained at TU Wien, Austria and the National Institute for Materials Science (NIMS), Japan. The power factor of the well-known Ni-Cu system is plotted for comparison. \textbf{d}, Composition-dependent figure of merit of Ni-Au and Ni-Cu binary alloy systems. Dashed lines are theoretical calculations from the DOS within the paramagnetic regime (details see Methods). Inset shows that the bandwidth of Ni 3$d$ states decreases for NiAu with respect to NiCu. The enhanced localization leads to a strongly increased thermoelectric performance of NiAu alloys.}
    \label{Fig4}
  \end{minipage}
\end{figure*}

Following this comprehensive assessment we experimentally studied the structural and TE properties of the binary Ni-Au system over the whole temperature (4\,--\,1100\,K) and composition range. 
In our endeavour we had to overcome the metastable nature of Ni$_{x}$Au$_{1-x}$. Its peculiar phase diagram involves a miscibility gap at low temperatures due to a large size mismatch of the lattice constants and a narrow region of solid solubility at high temperatures (see Extended Data Fig.\,4). Nevertheless, we succeeded in obtaining the desired \textit{fcc} phase at ambient temperatures by rapidly quenching NiAu samples in cold water, realizing record-high $PF$ up to 34\,mWm$^{-1}$K$^{-2}$ (Fig.\,\ref{Fig4}c). As seen in Fig.\,\ref{Fig4}a, a heating-cooling cycle of a sample with Ni ($x=0.37$) reestablishes the mixed phase with Ni-rich and Au-rich regions, evident from a reduction of $|S|$. To evade the phase-segregated regime, we studied the single-phase solid solution at $T\geq 1000$~K below the melting point, where we achieve the highest $zT\approx 0.45$ (see Fig.\,\ref{Fig4}d). While the peak values are realized at $\approx 10\,$at.\% Ni for $PF$ and $\approx 40$\,at.\% Ni for $zT$, the TE performance surpasses that of constantan (NiCu) by several times over the entire range of compositions -- a manifestation of the different bandwidth of Ni atoms in Cu or Au lattices.

A closer assessment of $S(T)$ in Fig.\,\ref{Fig4}a yields a linearly increasing Seebeck coefficient that saturates at high temperatures for all NiAu alloys, in good agreement with our theoretical estimation in Fig.\,\ref{Fig4}b. Note the quantitative agreement  with experimental data at $x\approx 0.4$, where the electron-hole asymmetry of $s$--$d$ scattering is most pronounced; deviations between experiment and theory become larger once inelastic phonon scattering becomes dominant away from the optimum TE performance. We further notice that alloys with $x<0.3$ do not show hysteretic behavior of $S(T)$ upon repeated thermal cycles, indicating that these are stable compositions that do not dissociate into separate phases (see Extended Data Fig.\,7). 
As elaborated in Methods, thermoelectric measurements were independently carried out on numerous samples in different laboratories (TU Wien, NIMS), confirming the validity and reproducibility of our experimental data.

The main finding of our study are the record-high power factors in Ni$_{x}$Au$_{1-x}$, reaching peak values of $34\,$mWm$^{-1}$K$^{-2}$ at 560\,K and $29\,$mWm$^{-1}$K$^{-2}$ at room temperature.
For $x=0.1$, we find a colossal average $PF\approx 30\,$mWm$^{-1}$K$^{-2}$ over an extremely broad temperature range 300--1100\,K. Such ultrahigh values above room temperature exceed those of state-of-the-art TE semiconductors by an order of magnitude and are even much larger than the giant power factors in strongly correlated $f$ electron systems like \ce{YbAl3} \cite{rowe2002electrical} and \ce{CePd3} \cite{gambino1973anomalously} or the magnon-drag metal cobalt \cite{watzman2016magnon}. This implies a great potential for electrical power generation, especially if the heat source is large \cite{liu2015n}, although the high cost of Au poses constraints on widespread use in the industry. Another potential application for materials with high $PF$ is the field of active cooling, where the Peltier effect is used to pump heat from the hot side towards ambient temperature. It was shown that in such cases the relevant quantity to be maximized is the so-called effective thermal conductivity $\kappa_\text{eff}=\kappa + PF\,\,T_\text{H}^2/(2\,\Delta T)$, with the second term being the active cooling contribution \cite{zebarjadi2015electronic}. Recent works showed that elemental Co \cite{adams2019active} ($\kappa_\text{eff}=780$\,Wm$^{-1}$K$^{-1}$) and carbon nanotube fibers\cite{komatsu2021macroscopic} (1190\,Wm$^{-1}$K$^{-1}$) can exceed the thermal conductivity of pure copper for $\Delta T =1\,$K \cite{adams2019active}. Under the same conditions, we obtain even higher $\kappa_\text{eff}\approx 1400$~Wm$^{-1}$K$^{-1}$ for \ce{Ni_{0.1}Au_{0.9}}, significantly exceeding $\kappa_\text{eff}$ of any material reported so far. Due to their great mechanical properties (high ductility, flexibility) and since they are very easy to manufacture and produce, these alloys are ideal candidates for active cooling applications, \textit{e.g.}, in integrated circuits.
 
Apart from niche applications such as thermocouples, metals have been discarded as TE materials, due to their usually small Seebeck effect. However, by using the concept of $s$--$d$ scattering, we have constructed a roadmap that can easily and accurately predict high TE performance in metallic alloys. Moreover, we have experimentally confirmed this new approach in binary NiAu alloys as a proof-of-concept and discovered unprecedentedly high power factors as well as the largest $zT$ ever reported for a metal, both over a broad temperature range. We emphasize that the enhancement principle is not bound to the specific elements Ni or Au, but is rather a general concept. Theoretical investigations reveal that the performance of metallic alloys can be significantly enhanced by bandwidth tuning via chemical pressure as demonstrated here. Indeed, our theoretical predictions indicate that even in pure Ni, an increase of the lattice parameter by $\approx 16$\,\% would result in a dramatic enhancement of the Seebeck effect, potentially reaching $zT>1$ at room temperature (see Extended Data Fig.\,3). Moreover, guided by our \textit{periodic table of densities of states}, we identify other binary systems with similar or even more promising electronic structures such as Ni-Na, Ir-Na, Rh-Na, Rh-K, Ir-K, Rh-Rb, Ir-Rb, Co-Si, Co-Ge, Co-Sn, Ni-In, Co-In, Sc-Cu, Sc-Ag, Sc-Au. Especially systems comprising transition and alkali metals show ultralocalized states next to $E_\text{F}$ due to the much larger atomic radii of the alkalis leading to very narrow $d$ bands upon lattice expansion. In conjunction with the broad $s$-type conduction bands this implies a great potential for strong \textit{s--d} scattering.

Although the $zT$ values of state-of-the-art inorganic semiconductors are still several times greater than those reported here, our work presents a major breakthrough in the search for high-performance metallic thermoelectrics and sets the stage for new avenues of TE research utilizing inter-orbital scattering. Metallic alloys with high TE performance have many advantages compared to their semiconducting counterparts, which often face problems such as thermal and mechanical degradation at the interface between the TE semiconductor and metal electrodes, impeding widespread application of TE devices \cite{yan2022high}. Due to their chemical inertness, mechanical strength, very high ductility and superior processability, metallic alloys have a huge potential to expand the field of thermoelectric applications, especially if high-performance systems consisting of cheap and abundant elements are realized.

\vspace{0.0cm}
\bibliographystyle{apsrev4-2}
\renewcommand{\bibsection}{}

\clearpage
\section*{\textsf{\fontsize{14}{14}\selectfont{Methods}}}
\noindent \textbf{Electronic structure calculations.} \textit{Ab initio} electronic structure calculations of binary metallic alloys were performed within the framework of density functional theory making use of self-consistent bulk Green-function formalism on the basis of the coherent potential approximation \cite{abrikosov1993self,ruban1999calculated}. To treat exchange correlation effects the Perdew-Wang parametrization of the local density approximation was used \cite{perdew1992accurate}. A dense grid of $75\times 75\times 75$ $k$-points was used in the calculation, corresponding to 9880 $k$-points in the irreducible Brillouin zone. Electronic densities of states of the pure transition metal elements were partly taken from the Materials Project database and partly calculated using the Vienna Ab Initio Simulation Package VASP with similar computational parameters.\\

\noindent \textbf{Transport modelling.} The total Seebeck coefficient of a multi-band electronic conductor results by weighting the respective contributions of the Seebeck coefficient with the electrical conductivities. Thus, in the case of \ce{Ni_{x}Au_{1-x}} alloys with primarily two types of charge carriers, $s$ electrons with high mobility and less mobile $d$ electrons, the net diffusion thermopower can be written as

\begin{equation}
S_\text{tot}=\frac{S_1\sigma_1+S_2\sigma_2}{\sigma_1+\sigma_2}.
\end{equation}

\noindent Here, the indices $i=1,2$ correspond to the $s$- and $d$-like electrons, respectively. The energy-dependent conductivity is given by $\sigma(E)\propto N(E)\,\tau(E)$, with $N(E)$ and $\tau(E)$ being the electronic density of states and relaxation time, respectively. For $s$-like conduction electrons scattering into empty $d$ states ($s-d$ scattering) the electrical conductivity yields $\sigma_1 \propto N_s/N_d$ since the scattering rate $\tau^{-1} \propto N_d$, as given by Fermi's golden rule. On the other hand, for the scattering mechanism where $d$-like electrons scatter into $d$ states ($d-d$ scattering), the electrical conductivity does not depend on the density of states, $\sigma_2 \propto N_d/N_d$. Using the Mott expression, $S\propto -\partial\, \text{ln}\,\sigma/\partial E$, as well as $\sigma_1>>\sigma_2$, Eq.\,1 from above can be rewritten to a first approximation

\begin{equation}
S_\text{tot} \propto \frac{-\frac{\partial\, \text{ln}\, \sigma_1}{\partial E}\sigma_1-\frac{\partial\, \text{ln}\, \sigma_2}{\partial E}\sigma_2}{\sigma_1+\sigma_2} \approx -\frac{\partial\, \text{ln}\, \sigma_1}{\partial E}.
\end{equation} 

\noindent Consequently, the Seebeck coefficient is given by the \textit{positive} logarithmic derivative of the density of states of the $d$-like electrons

\begin{equation}
S=+\frac{\pi^2k_\text{B}^2T}{3e}\left(\frac{\partial\,\text{ln}\,N}{\partial E}\right)_{E\approx E_\text{F}}.
\end{equation}

\noindent With $zT=S^2\sigma T/(L\sigma T+\kappa_\text{ph})\approx S^2/L$, this yields a simple formula for the figure of merit. To conduct a more in-depth examination of the thermoelectric properties at high temperatures, where the Fermi-Dirac distribution is broadened, the transport integrals must be solved

\begin{equation}
S(T)=\frac{k_\text{B}}{e}\dfrac{\int_{-\infty}^{\infty}\sigma(E)\frac{E-\mu}{T}\frac{-\partial f_0}{\partial E}dE}{\int_{-\infty}^{\infty}\sigma(E)\frac{-\partial f_0}{\partial E}dE}.
\end{equation}

\noindent Here, $f_0(E,\mu,T)$ represents the Fermi-Dirac distribution function, $\mu$ the chemical potential and $\sigma(E)\propto 1/N_d(E)$ the transport function. In order to theoretically estimate the composition-dependent figure of merit of binary metallic alloys comprising transition metals from group 10 and group 11 elements, the Seebeck coefficient of $s$--$d$ scattering was calculated via Eq.\,4. The $s$--$d$ scattering in $X_xY_{1-x}$ alloys is mainly caused by the elastic impurity scattering at the $X$ atoms. However, it is important to consider the various scattering processes when calculating the Seebeck coefficient, particularly at high temperatures where inelastic contributions like phonon scattering become more significant. To account for inelastic contributions at high temperatures the Nordheim-Gorther rule was used 

\begin{equation}
S_\text{tot}=\frac{S_{s-d}\rho_{s-d}+S_\text{in}\rho_\text{in}}{\rho_{s-d}+\rho_\text{in}}\approx \frac{S_{s-d}\rho_{s-d}}{\rho_{s-d}+\rho_\text{in}}.
\end{equation}

\noindent Here, $S_{s-d}$ represents the Seebeck coefficient due to $s$--$d$ scattering which was calculated by solving the transport integrals shown in Eq.\,4 using $\sigma(E)\propto 1/N_d(E)$. $\rho_{s-d}$ is the resistivity due to $s$--$d$ scattering and $S_\text{in},\,\rho_\text{in}$ are the inelastic contributions. In our model, we assumed $S_{s-d}>> S_\text{in}$, $\rho_{s-d}=\rho_0$, and hence, $\rho_\text{in}=\rho_\text{tot}-\rho_0$, due to Mathiessen's rule, with $\rho_0$ being the residual resistivity at low temperatures arising from elastic impurity scattering. $\rho_0$ and $\rho_\text{tot}$ are obtained from experiments (for Ni-Ag, where no experimental data exist, the resistivity values of the Ni-Cu system were taken). Within this framework, $zT\propto S^2$ was calculated without any free parameters, yielding very good qualitative and quantitative agreement with experimental high-temperature data. The Lorenz number $L$ was estimated from a simple formula $L=1.5+\text{exp}(-|S|/116)$, which is widely used for thermoelectric materials \cite{kim2015characterization}.

\section*{\large Experimental}
\noindent \textbf{Sample preparation.}
Binary NiAu alloys were prepared by stoichiometrically weighing 99.95\,\% pure Ni and 99.99\,\% pure Au bulk pieces. The pure metals were mounted in a copper hearth under argon atmosphere and melted by high-frequency induction heating. The as-cast ingots were cut using a diamond cutting wheel. Due to the metastable nature of the Ni-Au phase diagram, the properties of NiAu alloys depend strongly on the synthesis procedure. Samples which are slowly cooled from the melt dissociate into two separate phases, one gold-rich phase and one of almost pure Ni, resulting in two-phase microstructures (see Extended Data Fig.\,4). Therefore, structural characterization (x-ray diffraction, scanning electron microscopy, electron dispersive x-ray analysis) were employed to study the microstructure and phase homogeneity of our samples.
\newpage
\noindent \textbf{Sample characterisation.}
\subsection*{i) X-ray diffraction.}
\noindent The crystal structure and phases of all samples were analyzed via x-ray diffraction in a Bragg-Brentano geometry using conventional Cu-K$\alpha$ radiation. Slowly-cooled samples displayed two $fcc$ phases, whereas quenched samples showed a single $fcc$ phase. Quenched samples were measured from different sides (bottom side and top side) to ensure no spatial inhomogeneities were present within the samples. According to the Ni-Au phase diagram, there exists a phase boundary towards a single-phase solid solution at high temperatures at all compositions. To verify that our NiAu samples become single-phase at high temperatures within the time scale of the high-temperature thermoelectric property measurement, temperature-dependent \textit{in situ} x-ray diffraction experiments have been performed from room temperature up to 1273\,K, sufficiently above the phase transition temperature (see Extended Data Fig.\,4). We confirmed that within $< 30\,$ min, even the slowly cooled two-phase sample with the largest amount of secondary phase, transforms into a single $fcc$ phase.
\subsection*{ii) Microstructure and composition.}
\noindent The microstructure and composition were investigated by using a scanning electron microscope (SEM) and energy-dispersive x-ray analysis (EDX) at the USTEM, TU Wien. Slowly-cooled samples displayed a two-phase microstructure as can be seen in Extended Data Fig.\,4. However, macroscopically, the composition as well as the microstructure were confirmed to be consistent and homogeneous across the sample. Hence, this allowed us to use these samples for rapid thermal quenching in water, which resulted in homogeneous single-phase NiAu alloys of the desired nominal composition.\\

\noindent \textbf{Property measurements.}
Thermoelectric transport measurements have been performed at different setups at TU Wien in Austria as well as at the National Institute for Materials Science in Japan.
\subsection*{i) Low-temperature transport.}
\noindent Low-temperature investigations (4--300\,K) of the electrical resistivity and Seebeck coefficient have been performed at our in-house setups at the Institute of Solid State Physics, TU Wien. The resistivity was measured using a four-probe method with an ac resistance bridge (Lakeshore) and an excitation current of 31.6 mA. Thin gold wires were spot-welded onto the sample surface in the appropriate geometry and the sample was mounted in a Helium bath cryostate. The low-temperature Seebeck coefficient was measured by making use of a toggled heating technique with two constantan--chromel thermocouples that were thermally contacted to the sample.
\subsection*{ii) High-temperature transport.}
\noindent High-temperature measurements of the Seebeck coefficient and electrical resistivity have been performed by making use of a commercially available setup (ZEM3 by ULVAC) at TU Wien as well as at NIMS. Additionally, the thermal conductivity was measured for one representative sample (33\,at.\% Ni) at TU Wien using a LightFlash (LFA 500) diffusivity measurement setup by Linseis. This allowed us to confirm that the thermal conductivity of NiAu alloys is entirely dominated by electronic heat transport $\kappa \approx \kappa_\text{el}$. Indeed, due to the very soft lattice and low Debye temperature of Au ($\Theta_\text{D}\approx 165\,$K), the lattice contribution of the thermal conductivity makes up for less than 1\,\% of $\kappa$ in pure Au. In concentrated Au-rich CuAu alloys, which are struturally similar to the NiAu alloys studied here, $\kappa_\text{ph}$ is only around $1$\,Wm$^{-1}$K$^{-1}$ at room temperature and $0.5$\,Wm$^{-1}$K$^{-1}$ at 1100\,K \cite{ho1978thermal}. Calculating $\kappa_\text{el}$ by using an estimate for the Lorenz number as derived in Ref.\,\cite{kim2015characterization} reveals almost perfect agreement with the experimental thermal conductivity ($\kappa_\text{exp}\approx 27$, $\kappa_\text{el}\approx 28$\,Wm$^{-1}$K$^{-1}$ at 300\,K).

\section*{\large Data availability}
\noindent The data are available from the
corresponding author upon reasonable request.
\section*{\large Code availability}
\noindent The computer codes are available
from the corresponding author upon reasonable request.

\vspace*{0.3cm}
\footnotesize{
\noindent \textbf{Acknowledgements} Financial support for F.G., M.P., A.R., T.M. and E.B. came from the Japan Science and Technology Agency (JST), program MIRAI, JPMJMI19A1. We acknowledge the X-ray Center at TU Wien and Werner Artner for assisting with the high-temperature x-Ray diffraction experiments as well as providing their equipment. Furthermore, USTEM at TU Wien is acknowledged for providing the scanning electron microscope to study the microstructure and composition of NiAu alloys.}

\vspace*{0.3cm}
\footnotesize{
\noindent \textbf{Author contributions} F.G., M.P. and A.P. conceptualised the work and planned out the outline of the draft. S.K. performed density functional theory calculations of alloy-averaged densities of states. A.R. and F.G. analyzed the composition, microstructure and crystal structure of NiAu alloys. F.G. synthesized the samples and performed thermoelectric transport measurements from 4 to 860\,K at TU Wien. C.B. performed thermoelectric transport measurements from 300 to 1130\,K at NIMS, Japan. T.M., E.B. and A.P. supervised the work and organized the funding. All authors discussed the work and modified the manuscript.}

\vspace*{0.3cm}
\footnotesize{
\noindent \textbf{Competing interests} The authors declare no competing interest.}

\clearpage
\captionsetup[figure]{name=Extended Data Fig. ,labelsep=space}
\renewcommand{\figurename}{Extended Data Fig.}
\setcounter{figure}{0}

 \begin{figure*}[h!]
\newcommand{\setwidth}{0.45}
			\centering
			\hspace*{0cm}
		\includegraphics[width=1\textwidth]{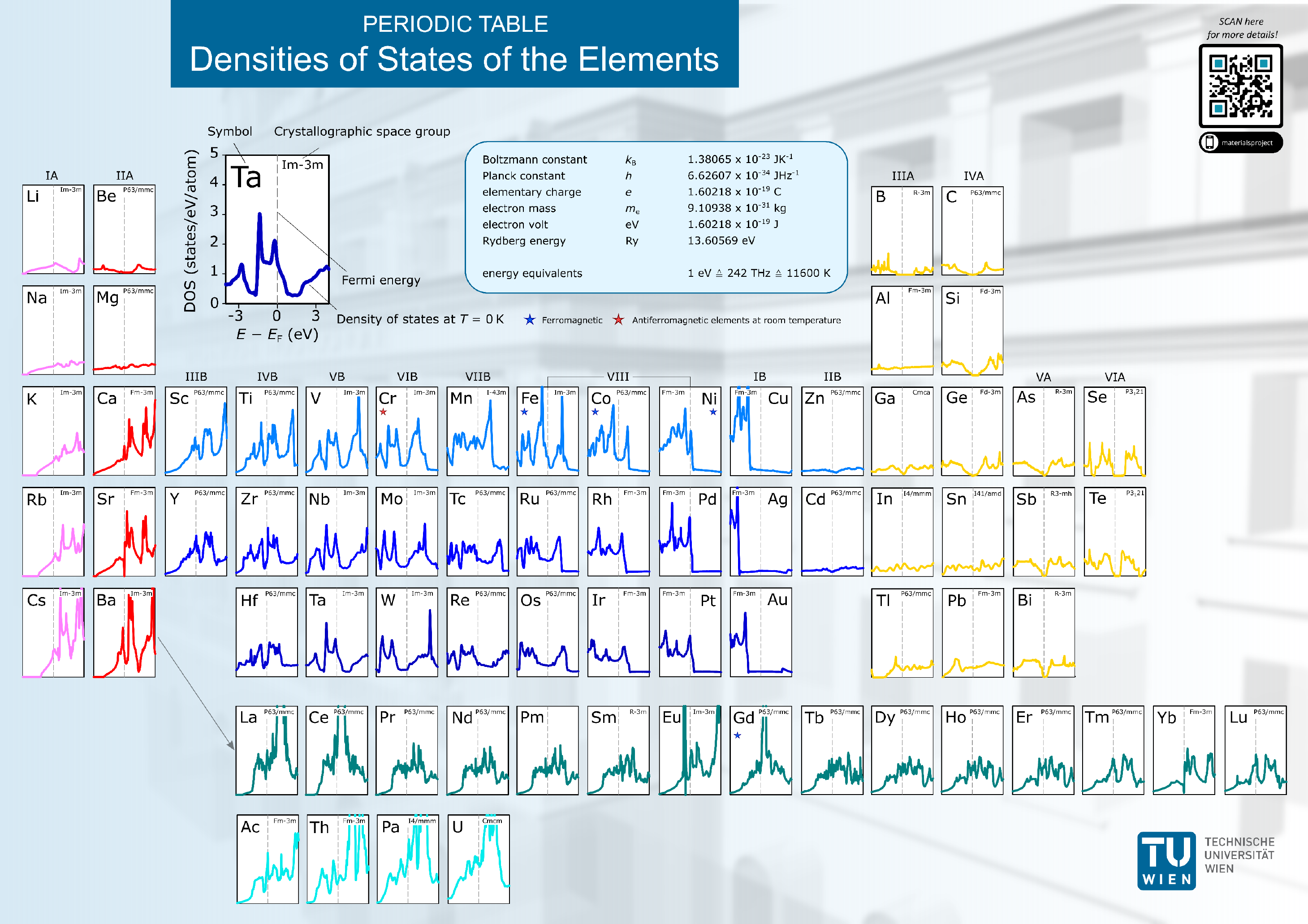}
	\caption{\textbf{$\vert$ Periodic table of densities of states of the elements.} Densities of states as a function of energy near the Fermi level for the different elements of the periodic table. Computational data were taken from the Materials Project database for the experimentally observed room-temperature crystal structure under ambient conditions.} 
	\label{ExtendedFig1}
\end{figure*}
\clearpage

\begin{figure*}[t!]
	\newcommand{\setwidth}{0.45}
			\centering
			\hspace*{0cm}
			\includegraphics[width=0.9\textwidth]{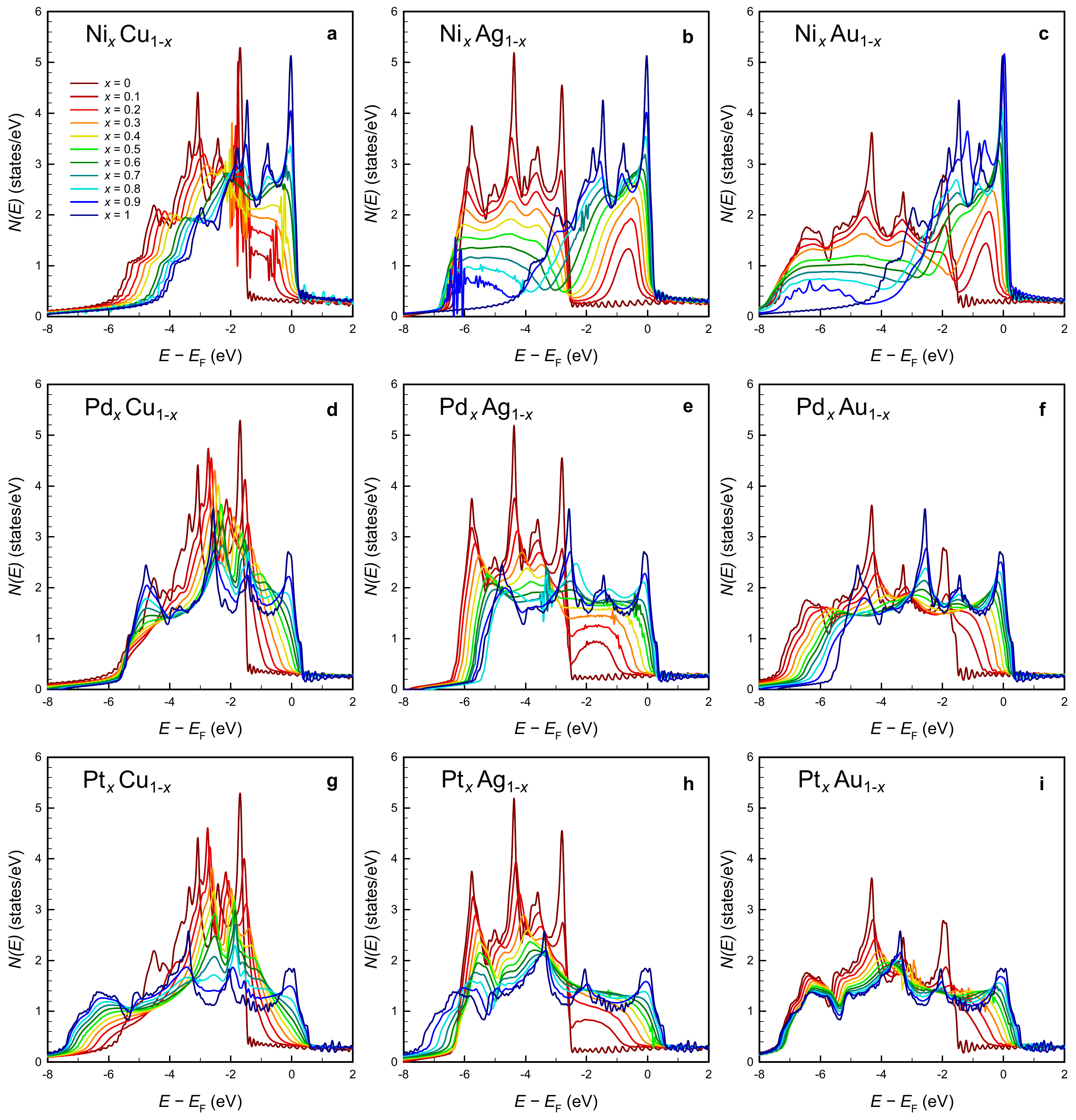}
	\caption{\textbf{$\vert$ Densities of states of binary metallic alloys.} Alloy-averaged densities of states of binary alloys from transition metals of group 10 with transition metals of group 11 elements; calculated for different alloy concentrations.} 
	\label{ExtendedFig2}
\end{figure*}

\begin{figure*}[t!]
	\newcommand{\setwidth}{0.45}
			\centering
			\hspace*{-0.5cm}
			\includegraphics[width=0.9\textwidth]{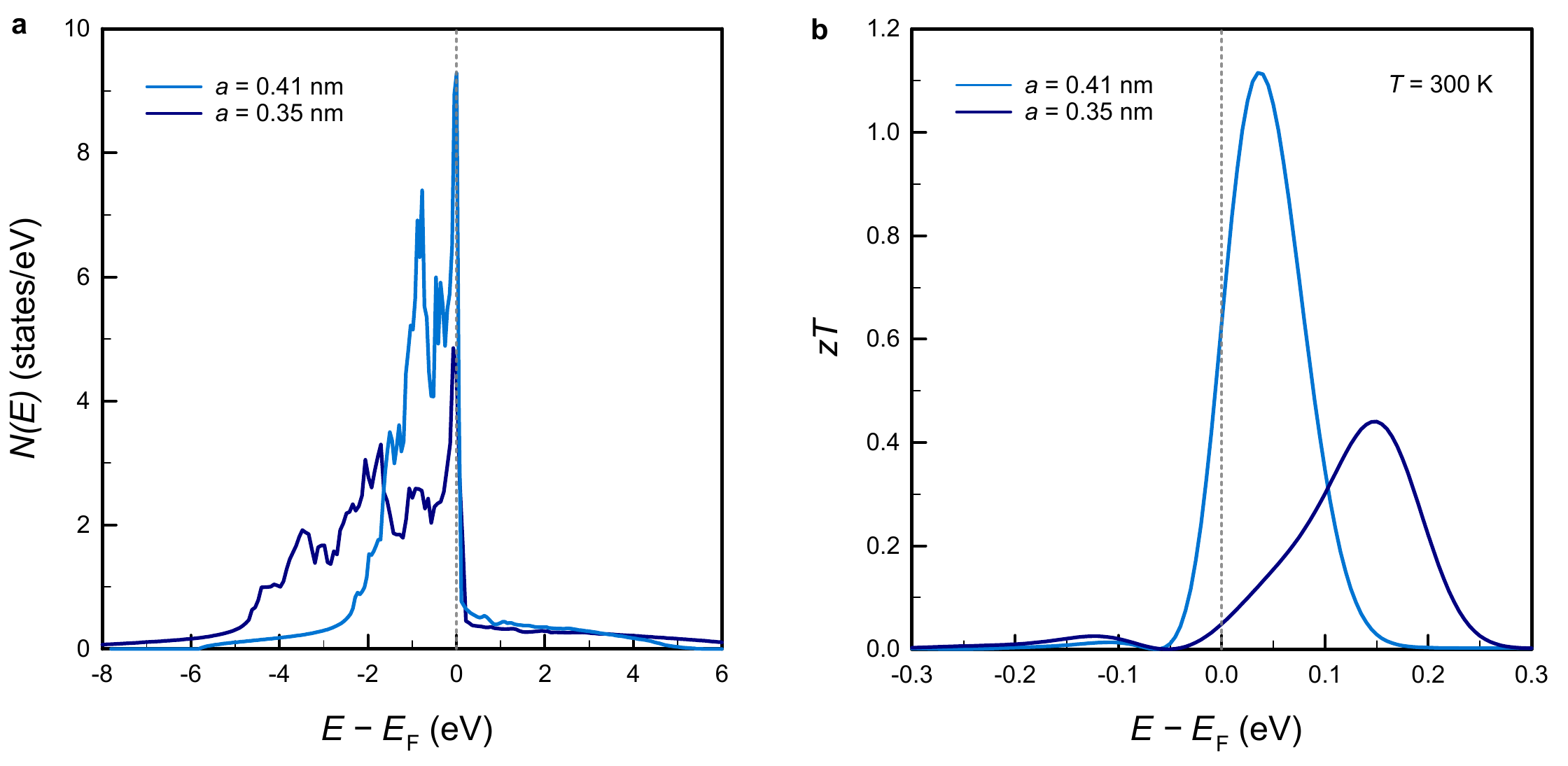}
	\caption{\textbf{$\vert$ Bandwidth-dependent thermoelectric performance of pure nickel.} \textbf{a}, Densities of states of Ni, calculated for different lattice parameters (i) the experimental lattice parameter of Ni and (ii) for a fictitious Ni crystal with a $\approx 16\,\%$ larger lattice parameter (same as the one of pure Au). The localization of the Ni 3$d$ states increases as the bandwidth decreases with increasing lattice parameter. \textbf{b}, Dimensionsless figure of merit $zT = S^2/L$ corresponding to the densities of states in the left panel; calculated at room temperature by solving the transport integrals for different positions of the Fermi energy.} 
	\label{S3}
\end{figure*}
\clearpage

\begin{figure*}[t!]
	\newcommand{\setwidth}{0.45}
			\centering
			\hspace*{0cm}
			\includegraphics[width=1\textwidth]{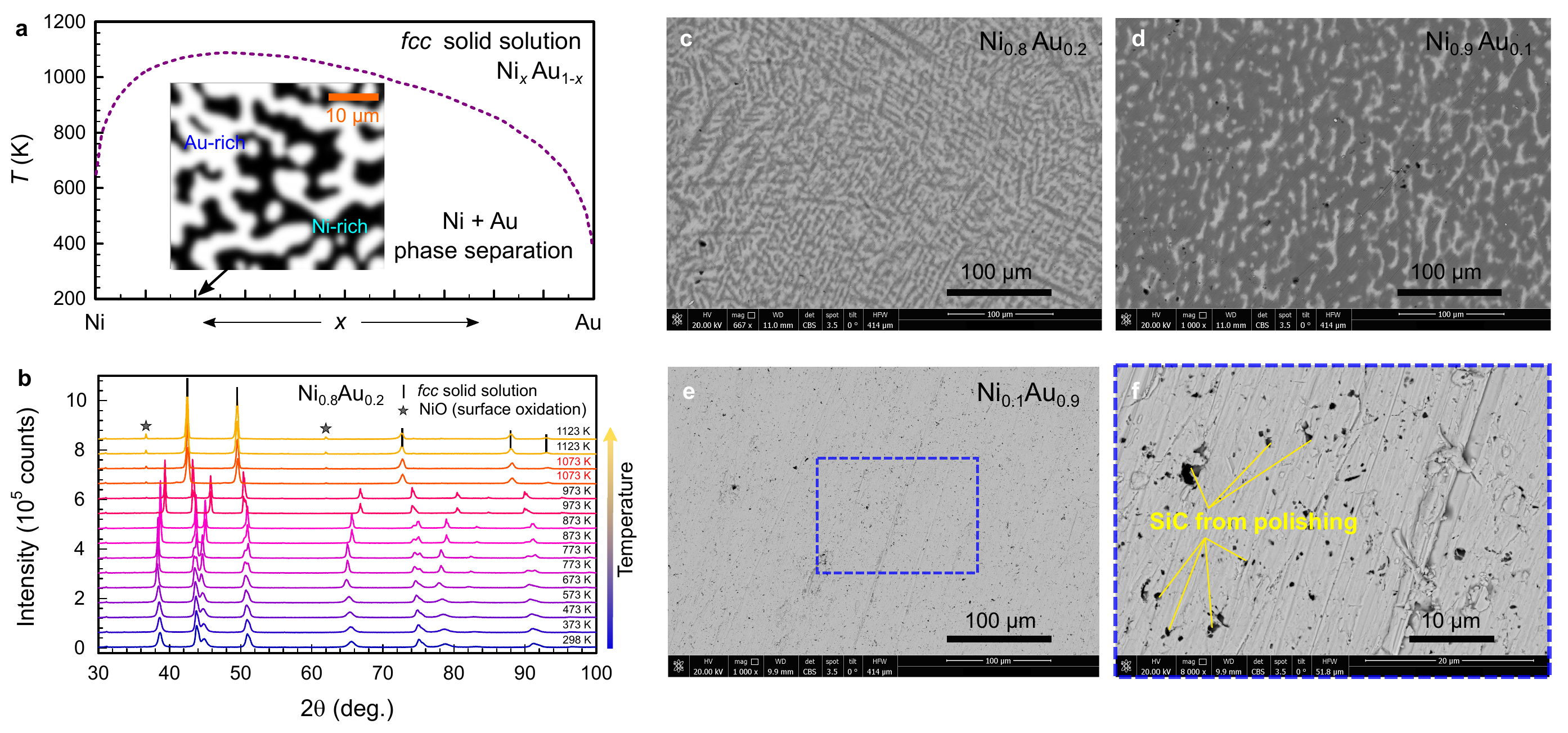}
	\caption{\textbf{Phase diagram and structural properties of the Ni-Au system.} \textbf{a}, Phase diagram of the binary Ni-Au system, showing a large miscibility gap and narrow region of solid solubility at high temperatures \cite{wang2005thermodynamic}. Inset shows a high-contrast scanning electron microscopy image for a slowly cooled two-phase NiAu sample. \textbf{b}, X-ray diffraction patterns at various temperatures for a two-phase \ce{Ni_{0.8}Au_{0.2}} sample, which transitions into a single-phase $fcc$ alloy at high temperatures. Each diffraction pattern was obtained within a time scale of approximately 30 minutes. \textbf{c-d}, Two-phase microstructure of slowly-cooled Ni-rich alloys. \textbf{e}, Single-phase microstructure of a Au-rich \ce{Ni_{0.1}Au_{0.9}} alloy. \textbf{f}, Holes and SiC inclusions from the cutting and polishing of the sample due to the soft and ductile nature of Au-rich alloys.} 
	\label{S4}
\end{figure*}
\clearpage

\begin{figure*}[t!]
	\newcommand{\setwidth}{0.45}
			\centering
			\hspace*{-0.5cm}
			\includegraphics[width=0.8\textwidth]{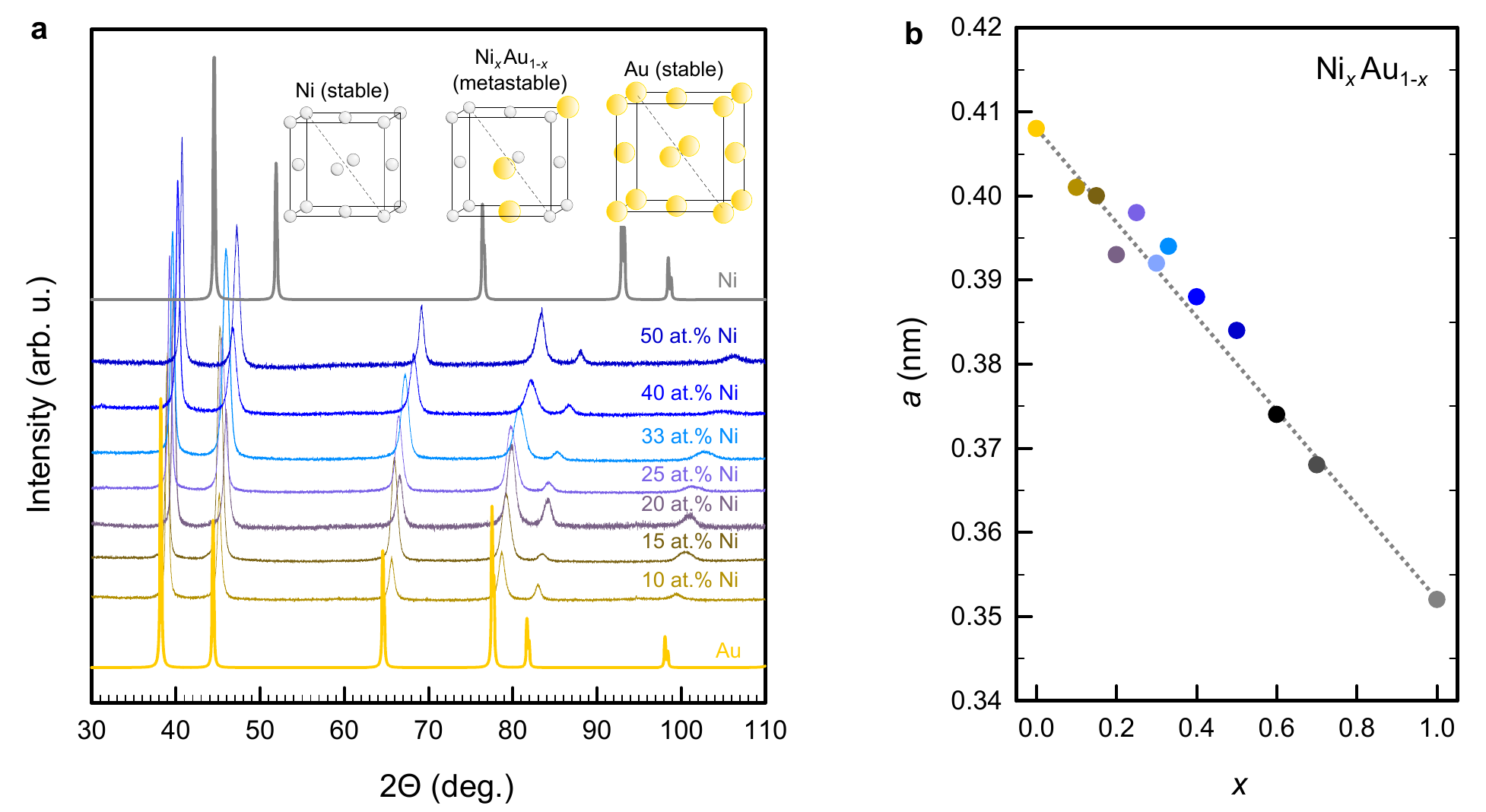}
	\caption{\textbf{$\vert$ Crystal structure of rapidly cooled NiAu alloys.} \textbf{a}, X-ray diffraction patterns of some selected NiAu alloys, which were rapidly cooled from the melt to retain a single $fcc$ phase \textbf{b}, Lattice parameters of the rapidly cooled Ni-Au alloy system as a function of Ni concentration.} 
	\label{S5}
\end{figure*}
\clearpage

\begin{figure*}[t!]
	\newcommand{\setwidth}{0.45}
			\centering
			\hspace*{-0.5cm}
			\includegraphics[width=1.05\textwidth]{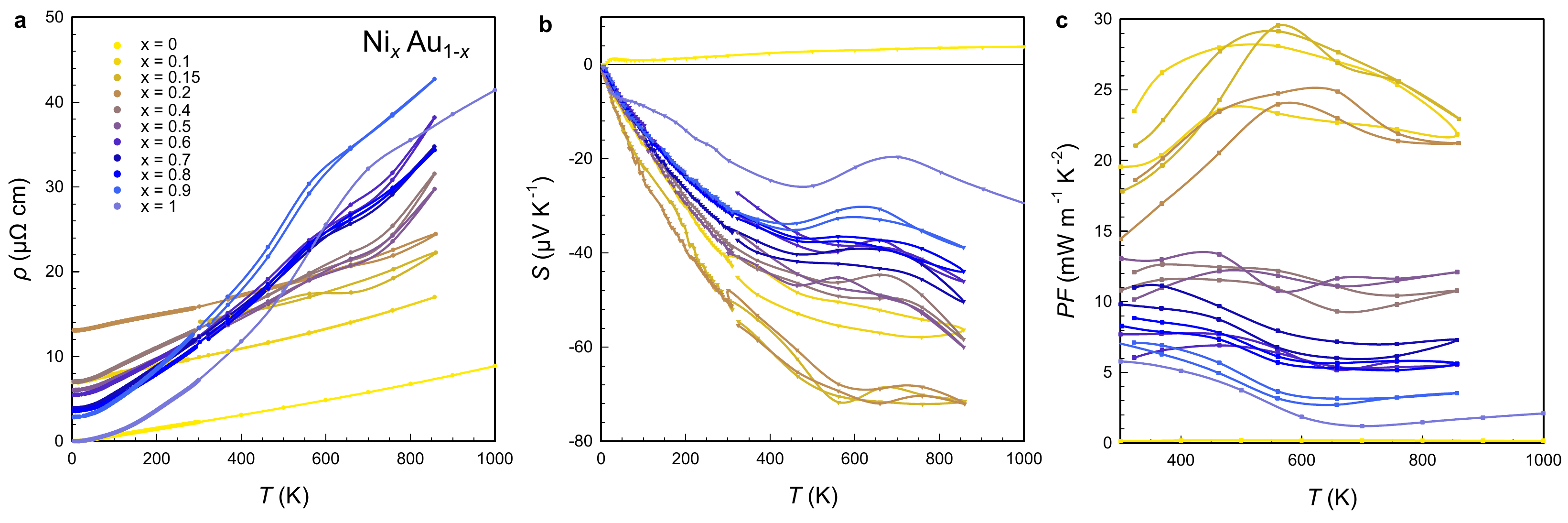}
	\caption{\textbf{$\vert$ Thermoelectric properties of slowly cooled NiAu alloys.} \textbf{a}, Electrical resistivity \textbf{b}, Seebeck coefficient \textbf{c}, power factor as a function of temperature. Slowly cooled \ce{Ni_xAu_{1-x}} alloys with $x\geq 0.4$ display a two-phase microstructure consisting of a Au-rich phase and a Ni-rich phase of almost pure Ni.} 
	\label{S6}
\end{figure*}
\clearpage

\begin{figure*}[t!]
	\newcommand{\setwidth}{0.45}
			\centering
			\hspace*{-0.5cm}
			\includegraphics[width=0.8\textwidth]{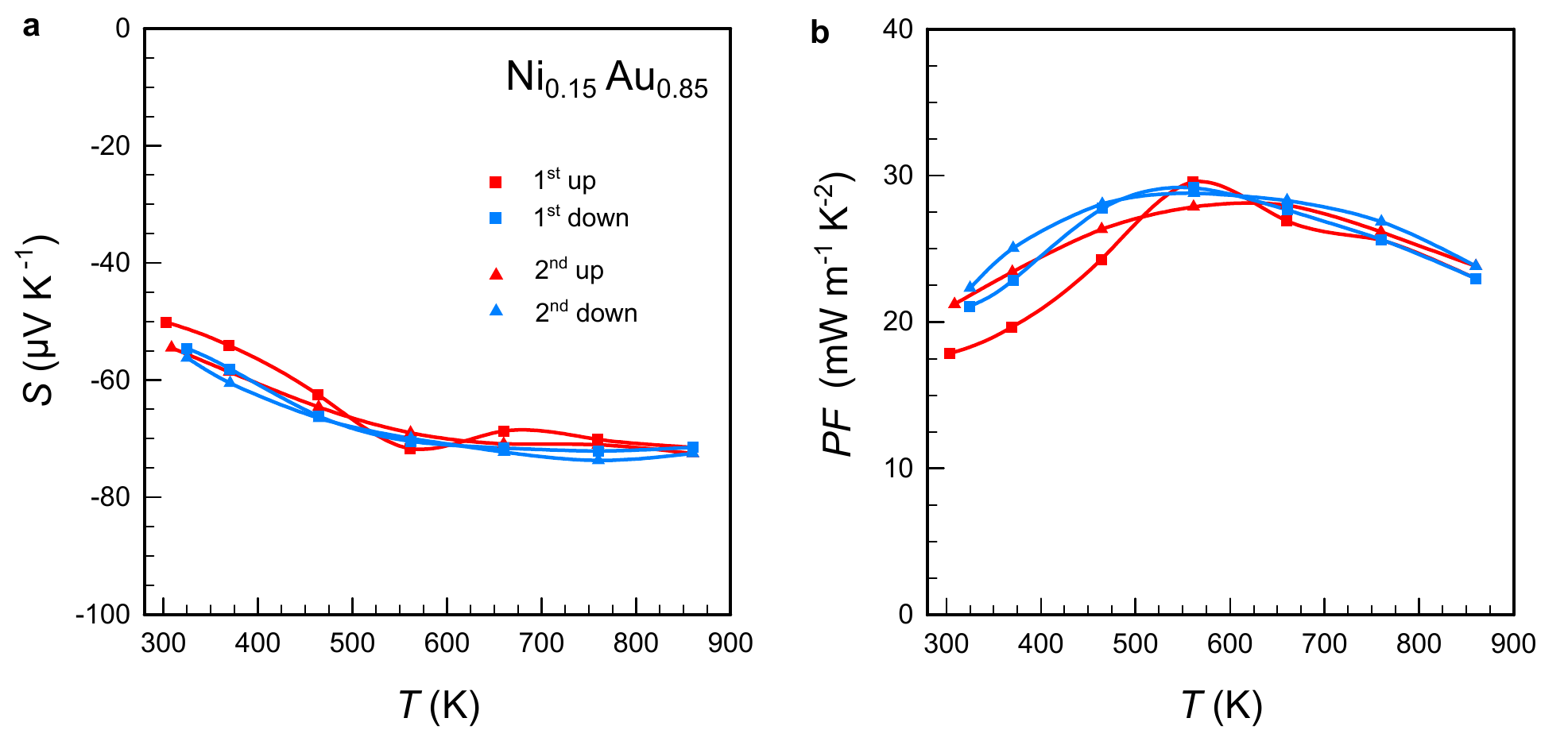}
	\caption{\textbf{$\vert$ Thermal stability and reproducibility of the thermoelectric properties of Au-rich NiAu alloy.} \textbf{a}, Seebeck coefficient \textbf{b}, power factor of \ce{Ni_{0.15}Au_{0.85}} as a function of temperature for two consecutive measurements with two heating and two cooling cycles.} 
	\label{S7}
\end{figure*}
\clearpage

\end{document}